\documentclass{iau}

\usepackage{amssymb}
\usepackage{amsmath}
\usepackage{mathtools}

\usepackage{graphicx}
\usepackage{subfigure}
\usepackage{color}
\usepackage{booktabs}
\usepackage[merge]{natbib}
\usepackage{grffile}
\usepackage{hyperref}
\hypersetup{breaklinks,colorlinks,citecolor=blue}

\newcommand{\eqn}[1]{(#1)}

\newcommand{\spcend}{\ensuremath{\:}}
\newcommand{\img}{\ensuremath{{\rm i}}}
\newcommand{\cconj}{\ensuremath{\ast}} 
 
\newcommand{\reals}{\ensuremath{\mathbb{R}}}

\newcommand{\integers}{\ensuremath{\mathbb{Z}}}
\newcommand{\naturals}{\ensuremath{\mathbb{N}}}

\newcommand{\ltwo}{\ensuremath{\mathrm{L}^2}}
\newcommand{\sphere}{\ensuremath{{\mathbb{S}^2}}}
\newcommand{\sothree}{\ensuremath{{\mathrm{SO}(3)}}}

\newcommand{\dx}{\ensuremath{\mathrm{\,d}}}
\newcommand{\dmu}[1]{\ensuremath{\dx \Omega(#1)}}

\newcommand{\deul}[1]{\ensuremath{\dx \varrho(#1)}}

\newcommand{\innerp}[2]{\ensuremath{\langle {#1},\: {#2} \rangle}}

\newcommand{\sa}{\ensuremath{\omega}}
\newcommand{\saa}{\ensuremath{\theta}}
\newcommand{\sab}{\ensuremath{\varphi}}
\newcommand{\sas}{\ensuremath{\saa, \sab}}
\newcommand{\eul}{\ensuremath{\mathbf{\rho}}}
\newcommand{\euls}{\ensuremath{\eula, \eulb, \eulc}}
\newcommand{\eula}{\ensuremath{\alpha}}
\newcommand{\eulb}{\ensuremath{\beta}}
\newcommand{\eulc}{\ensuremath{\gamma}}

\newcommand{\el}{\ensuremath{\ell}}
\newcommand{\m}{\ensuremath{m}}

\newcommand{\spin}{\ensuremath{s}}

\renewcommand{\exp}[1]{\ensuremath{{\rm e}^{#1}}}

\newcommand{\shf}[2]{\ensuremath{Y_{#1#2}}}

\newcommand{\shc}[3]{\ensuremath{{#1}_{{#2}{#3}}}}

\newcommand{\sshf}[3]{\ensuremath{{{}_{#3} Y_{#1#2}}}}

\newcommand{\sshc}[4]{\ensuremath{{}_{#4} {#1}_{{#2}{#3}}}}

\newcommand{\rotarg}[1]{\ensuremath{\mathcal{R_{#1}}}}

\newcommand{\rotmatarg}[1]{\ensuremath{\mathbf{R}_{#1}}}

\newcommand{\fs}{\ensuremath{{}_\spin f}}

\newcommand{\wav}{\ensuremath{\psi}}

\newcommand{\wcoeff}{\ensuremath{W}}

\newcommand{\wscale}{\ensuremath{j}}
\newcommand{\wscalemax}{\ensuremath{J}}
\newcommand{\wscalemin}{\ensuremath{J_0}}

\newcommand{\wavker}{\ensuremath{\kappa}}
\newcommand{\wavsteer}{\ensuremath{\zeta}}

\newcommand{\conv}{\ensuremath{\star}}

\renewcommand{\eqn}[1]{Eqn.~(#1)}

\renewcommand{\wav}{\ensuremath{\Psi}}

\renewcommand{\exp}[1]{\ensuremath{{\rm exp}{#1}}}

\title[On spin scale-discretised wavelets on the sphere] 
{On spin scale-discretised wavelets on the sphere for the analysis of
  CMB polarisation}

\author[McEwen et al.]  
{Jason D.\ McEwen$^1$, Martin B\"uttner$^2$, Boris Leistedt$^2$,\\
  Hiranya V.\ Peiris$^2$, Pierre Vandergheynst$^3$ \and Yves Wiaux$^4$
}

\affiliation{$^1$Mullard Space Science Laboratory (MSSL), University
  College London (UCL), \\Surrey RH5 6NT, UK \\ email: {\tt jason.mcewen@ucl.ac.uk} \\[\affilskip]
  $^2$Department of Physics and Astronomy, University College London
  (UCL), \\
  London WC1E 6BT, UK \\email: {\tt \{martin.buettner.11,
    boris.leistedt.11, h.peiris\}@ucl.ac.uk}\\[\affilskip]
  $^3$Institute of Electrical Engineering, Ecole Polytechnique
  F{\'e}d{\'e}rale de Lausanne (EPFL),\\ CH-1015 Lausanne,
  Switzerland \\email: {\tt pierre.vandergheynst@epfl.ch}\\[\affilskip]
  $^4$Institute of Sensors, Signals \& Systems, Heriot-Watt
  University,\\ Edinburgh EH14 4AS, UK \\email: {\tt y.wiaux@hw.ac.uk}
}

\pubyear{2014}
\volume{306}  
\pagerange{119--126}
\jname{Statistical Challenges in 21st Century Cosmology}
\editors{A.F. Heavens, J.-L. Starck, A. Krone-Martins eds.}
\begin{document}

\maketitle

\begin{abstract}
  A new spin wavelet transform on the sphere is proposed to analyse
  the polarisation of the cosmic microwave background (CMB), a spin
  $\pm 2$ signal observed on the celestial sphere.  The scalar
  directional scale-discretised wavelet transform on the sphere is
  extended to analyse signals of arbitrary spin.  The resulting spin
  scale-discretised wavelet transform probes the directional intensity
  of spin signals.  A procedure is presented using this new spin
  wavelet transform to recover E- and B-mode signals from partial-sky
  observations of CMB polarisation.

\end{abstract}

\firstsection 

\section{Introduction}

The polarisation of the cosmic microwave background (CMB) is a
powerful probe of the physics of inflation \citep{spergel:1997} and
the reionisation history of the Universe \citep{zaldarriaga:1997b}.
Numerous experiments have now measured CMB polarisation (some of the
more recent include: 
\citealt{hanson:2013, naess:2014, bicep2:I:2014}).  Although the
Planck satellite also measured CMB polarisation, polarisation data
were not included in the Planck 2013 release \citep{planck2013-p01}
but are anticipated later this year.

Since different physical processes often exhibit different symmetries,
their signatures in observables like CMB polarisation may behave
differently under a parity transform.  CMB polarisation can be
separated into parity even and parity odd components, so called E-
and B-mode components, respectively \citep{zaldarriaga:1997}.  Density
perturbations in the early Universe provide no mechanism to generate
B-mode polarisation in the CMB, whereas gravitational waves can induce
both E- and B-mode components.  The detection of primordial B-mode
polarisation would thus provide evidence for gravitational waves and
would provide a powerful probe of the physics of inflation.  
%

In these proceedings we outline a new spin wavelet transform on the
sphere to analyse observations of CMB polarisation,
a spin $\pm 2$ signal observed on the celestial sphere. In addition, we
describe a simple technique based on this wavelet framework to
separate E- and B-mode CMB polarisation components from partial-sky
observations.  We present a preliminary discussion only; further
details of these methods, fast implementations, and a rigorous
evaluation of their performance will be given in a series of forthcoming
articles.

\section{Spin scale-discretised wavelets on the sphere}

Scalar wavelets on the sphere \citep[e.g.][]{antoine:1998,
  antoine:1999, baldi:2009, mcewen:2006:cswt2, mcewen:2006:fcswt,
  mcewen:2013:waveletsxv, marinucci:2008, narcowich:2006, starck:2006,
  wiaux:2005, wiaux:2005c, wiaux:2007:sdw, leistedt:s2let_axisym} have
proved an effective tool for analysing the temperature anisotropies of
the CMB \citep[e.g.][]{vielva:2004, vielva:2006, mcewen:2005:ng,
  mcewen:2006:ng, mcewen:2008:ng, mcewen:2006:bianchi,
  mcewen:2006:isw, mcewen:2007:isw2, pietrobon:2006, fay:2008,
  feeney:2011b, feeney:2011a, bobin:2013, planck2013-p06,
  planck2013-p09, planck2013-p09a, planck2013-p20}.  For a somewhat
dated review see \cite{mcewen:2006:review}.  Spin wavelets to analyse
the polarisation of the CMB have been constructed by
\citet{geller:2008} and \citet{starck:2009}.  However, a spin wavelet
transform on the sphere capable of probing the directional intensity
of signals does not yet exist.\footnote{Spin curvelets
  \citep{starck:2009} could be used for a directional analysis however
  these are constructed on the base pixels of Healpix
  \citep{gorski:2005} and so do not live naturally on the sphere.}  We
propose such a transform here by extending the directional
scale-discretised wavelet transform of \citet{wiaux:2007:sdw} to
signals of arbitrary spin on the sphere.

Spin scale-discretised wavelets ${}_\spin \wav^{(\wscale)} \in
\ltwo(\sphere)$ can be constructed on the sphere $\sphere$ in an
analogous manner to the scalar wavelet construction
\citep{wiaux:2007:sdw, leistedt:s2let_axisym, mcewen:2013:waveletsxv},
that is simply by defining the spin harmonic coefficients of the
wavelets in the factorised form:
\begin{equation}
  \label{eqn:wav_factorized}
  \sshc{\wav}{\el}{\m}{\spin}^{(\wscale)} \equiv
  \wavker^{(\wscale)}(\el) \: \shc{\wavsteer}{\el}{\m}
  \spcend,
\end{equation}
where $\sshc{\wav}{\el}{\m}{\spin}^{(\wscale)} = \innerp{{}_\spin
  \wav^{(\wscale)}}{\sshf{\el}{\m}{\spin}}$ are the spin
$\spin\in\integers$ spherical harmonic coefficients of the wavelets,
with $\sshf{\el}{\m}{\spin}$ denoting the spherical harmonic functions
and $\el \in \naturals_0$, $\m \in \integers$, such that $\vert \spin
\vert \leq \el$ and $\vert \m \vert \leq \el$.  The \emph{kernel}
$\wavker^{(\wscale)} \in \ltwo(\reals^{+})$ controls the angular
localisation of the wavelets, while their directional properties are
controlled by the \emph{directionality component} $\wavsteer \in
\ltwo(\sphere)$, with harmonic coefficients $\shc{\wavsteer}{\el}{\m}
= \innerp{\wavsteer}{\shf{\el}{\m}}$.  The wavelet scale $\wscale \in
\naturals_0$ encodes the angular localisation of $\wav^{(\wscale)}$.
The kernel and directionality component are defined as in the scalar
setting \citep{wiaux:2007:sdw,mcewen:2013:waveletsxv}.

The wavelet transform of a spin signal $\fs \in \ltwo(\sphere)$ on the
sphere is defined by the directional convolution of \fs\ with the
wavelet ${}_\spin \wav^{(\wscale)} \in \ltwo(\sphere)$.  The wavelet
coefficients $\wcoeff^{{}_\spin \wav^{(\wscale)}} \in \ltwo(\sothree)$
thus read
\begin{equation}
  \label{eqn:analysis}
  \wcoeff^{{}_\spin \wav^{(\wscale)}}(\eul) \equiv ( \fs \conv {}_\spin\wav^{(\wscale)}) (\eul)
  \equiv \innerp{\fs}{\rotarg{\eul} \: {}_\spin\wav^{(\wscale)}}
  = \int_\sphere \dmu{\sa} \fs(\sa) (\rotarg{\eul} \: {}_\spin\wav^{(\wscale)})^\cconj(\sa)
  \spcend ,
\end{equation}
where $\sa=(\sas) \in \sphere$ denotes spherical coordinates with
colatitude $\saa \in [0,\pi]$ and longitude $\sab \in [0,2\pi)$,
$\dmu{\sa} = \sin\saa \dx\saa \dx\sab$ is the usual rotation
invariant measure on the sphere, and $\cdot^\cconj$ denotes complex
conjugation.  
The rotation operator is defined by 
\begin{equation}
  (\rotarg{\eul} \: {}_\spin\wav^{(\wscale)}) \equiv {}_\spin\wav^{(\wscale)}(\rotmatarg{\eul}^{-1} \cdot \sa)
  \spcend ,
\end{equation}
where $\rotmatarg{\eul}$ is the three-dimensional rotation matrix
corresponding to $\rotarg{\eul}$.  Rotations are specified by elements
of the rotation group $\sothree$, parameterised by the Euler angles
$\eul=(\euls) \in \sothree$, with $\eula \in [0,2\pi)$, $\eulb \in
[0,\pi]$ and $\eulc \in [0,2\pi)$.  
Note that the wavelet coefficients are a scalar signal defined on the
rotation group $\sothree$.  The wavelet transform of
\eqn{\ref{eqn:analysis}} thus probes the directional intensity of the
signal of interest \fs.

Provided the wavelets satisfy an admissibility property analogous to
the scalar setting, the original signal can be synthesised exactly
from its wavelet coefficients by
\begin{equation}
  \label{eqn:synthesis}
  \fs(\sa) = 
  \sum_{\wscale=\wscalemin}^\wscalemax \int_\sothree \deul{\eul}
  \wcoeff^{{}_\spin \wav^\wscale}(\eul) (\rotarg{\eul} \: {}_\spin \wav^\wscale)(\sa)
  \spcend ,
\end{equation}
where $\deul{\eul} = \sin\eulb \dx\eula \dx\eulb \dx\eulc$ is the
usual invariant measure on \sothree\ and $\wscalemin$ and $\wscalemax$
are the minimum and maximum wavelet scales considered, respectively,
i.e.\ $\wscalemin \leq \wscale \leq \wscalemax$.  Throughout this
description we have neglected to include a scaling function, which must be
introduced to capture the low-frequency content of the analysed signal
\fs.

\section{E- and B-mode separation}

CMB experiments measure the scalar Stoke parameters $I,Q,U \in
\ltwo(\sphere)$, where $I$ encodes the intensity and $Q$ and $U$ the
linear polarisation of the incident CMB radiation (the circular
polarisation component of the four Stokes parameters $V \in
\ltwo(\sphere)$ is zero).  The linear polarisation signal that is
observed depends on the choice of local coordinate frame.  The
component $Q \pm \img U$ transforms under a rotation of the local
coordinate frame by $\chi \in [0, 2\pi)$ as $(Q \pm \img
U)^\prime(\sa) = \exp{( \mp \img 2 \chi)}(Q \pm \img U)(\sa)$ and is
thus a spin $\pm 2$ signal on the sphere \citep{zaldarriaga:1997}.
The quantity $Q \pm \img U$ can be decomposed into parity even and odd
components by
$  \tilde{E}(\sa) = -\frac{1}{2}\bigl[ \bar{\eth}^2(Q+\img
    U)(\sa) + {\eth}^2(Q-\img 
    U)(\sa) \bigr] 
$
and
$  \tilde{B}(\sa) = \frac{\img}{2}\bigl[
    \bar{\eth}^2(Q+\img U)(\sa) - {\eth}^2(Q-\img 
    U)(\sa) \bigr] 
$ 
respectively, where $\tilde{E},\tilde{B} \in \ltwo(\sphere)$ and
$\eth$ and $\bar{\eth}$ are spin raising and lowering operators,
respectively \citep{zaldarriaga:1997}.  Recovering E- and B-modes from
full-sky observations is relatively straightforward, however in
practice we observe the CMB over only part of the sky, since microwave
emissions from our Galaxy obscure our view.  A number of techniques
have been developed to recover E- and B-modes from $Q$ and $U$ maps
observed on the partial-sky \citep[e.g.][]{lewis:2002a, bunn:2003,
  kim:2011, bowyer:2011}.  Here we propose a simple alternative
approach using the spin scale-discretised wavelet transform described
above (a similar approach using needlets has been proposed by
\citet{geller:2008}, however there are some minor differences since
spin needlets yield spin and not scalar wavelet coefficients).

First, consider the wavelet coefficients of the observable $Q + \img U$
signal computed by a \emph{spin} wavelet transform:
$\wcoeff^{{}_2 \wav^{(\wscale)}}_{Q+\img U}(\eul) \equiv \innerp{Q +
  \img U}{\rotarg{\eul}\:{}_2 \wav^{(\wscale)}}$.  Second, consider
the wavelet coefficients of the unobservable $\tilde{E}$ and $\tilde{B}$
signals computed by a \emph{scalar} wavelet transform:
$\wcoeff^{{}_0 \tilde{\wav}^\wscale}_{\tilde{E}}(\eul) \equiv
\innerp{\tilde{E}}{\rotarg{\eul}\:{}_0 \tilde{\wav}^\wscale}$ and $
\wcoeff^{{}_0 \tilde{\wav}^\wscale}_{\tilde{B}}(\eul) \equiv
\innerp{\tilde{B}}{\rotarg{\eul}\:{}_0 \tilde{\wav}^\wscale}$.  If
the wavelet used in the scalar wavelet transform is a spin lowered
version of the wavelet used in the spin wavelet transform, i.e.\
${}_0 \tilde{\wav}^\wscale = \bar{\eth}^2 {}_2 \wav^\wscale$, then
the wavelet coefficients of $\tilde{E}$ and $\tilde{B}$ are simply
related to the wavelet coefficients of $Q + \img U$ by
$\wcoeff^{{}_0 \tilde{\wav}^\wscale}_{\tilde{E}}(\eul) = - {\rm Re}
\bigl[ \wcoeff^{{}_2 \wav^\wscale}_{Q+\img U} (\eul) \bigr]$ and
$\wcoeff^{{}_0 \tilde{\wav}^\wscale}_{\tilde{B}}(\eul) = - {\rm
  Im}\bigl[\wcoeff^{{}_2 \wav^\wscale}_{Q+\img U} (\eul)\bigr]$,
respectively.  

This leads to an elegant procedure to recover E- and B-modes from $Q$
and $U$ maps observed over the partial-sky.  Firstly, compute the spin
wavelet transform of $Q + \img U$.  Secondly, mitigate the impact of
the partial sky coverage in wavelet space, where signal content (and
thus the influence of the mask) is localised in scale and position
simultaneously.  Thirdly, reconstruct $\tilde{E}$ and $\tilde{B}$ maps
by inverse scalar wavelet transforms of the real and imaginary components,
respectively, of the processed spin wavelet coefficients.







\bibliographystyle{mymnras_eprint_noinitials}
\bibliography{bib_journal_names_short,bib_myname,bib}

\end{document}